\documentclass
[aps,prl,twocolumn,amsmath,amssymb,showpacs,superscriptaddress,notitlepage]{revtex4-1}%
\usepackage{amsmath,amssymb,amsfonts,bm}
\usepackage{graphicx}
\usepackage{epstopdf}
\usepackage{dcolumn}
\usepackage{mathrsfs}
\usepackage{bbold}
\usepackage{dsfont}
\usepackage{dcolumn}
\usepackage[colorlinks=true,linkcolor=blue,citecolor=blue,
urlcolor=blue,bookmarks=false]{hyperref}
\usepackage{amsmath}
\usepackage{amsfonts}
\usepackage{amssymb}%
\usepackage{float}
\usepackage{changes}

\begin{document}

\title{Nonlinear Hall effect on a disordered lattice}

\author{Rui Chen}
\affiliation{Department of Physics and Shenzhen Institute for Quantum Science and Engineering, Southern University of Science and Technology (SUSTech), Shenzhen 518055, China}
\affiliation{Department of Physics, Hubei University, Wuhan 430062, China}

\author{Z. Z. Du}
\affiliation{Department of Physics and Shenzhen Institute for Quantum Science and Engineering, Southern University of Science and Technology (SUSTech), Shenzhen 518055, China}
\affiliation{Shenzhen Key Laboratory of Quantum Science and Engineering, Shenzhen 518055, China}
\affiliation{International Quantum Academy, Shenzhen 518048, China}

\author{Hai-Peng Sun}
\affiliation{Institute for Theoretical Physics and Astrophysics, University of W\"urzburg, 97074 W\"urzburg, Germany\looseness=-1}

\author{Hai-Zhou Lu}
\email{Corresponding author: luhz@sustech.edu.cn}
\affiliation{Department of Physics and Shenzhen Institute for Quantum Science and Engineering, Southern University of Science and Technology (SUSTech), Shenzhen 518055, China}
\affiliation{Shenzhen Key Laboratory of Quantum Science and Engineering, Shenzhen 518055, China}
\affiliation{International Quantum Academy, Shenzhen 518048, China}
\affiliation{Quantum Science Center of Guangdong-Hong Kong-Macao Greater Bay Area (Guangdong), Shenzhen 518045, China}

\author{X. C. Xie}
\affiliation{International Center for Quantum Materials, School of Physics, Peking University, Beijing 100871, China}
\affiliation{Interdisciplinary Center for Theoretical Physics and Information Sciences (ICTPIS), Fudan University, Shanghai 200433, China}
\affiliation{Hefei National Laboratory, Hefei 230088, China}

\begin{abstract}
The nonlinear Hall effect has recently attracted significant interest due to its potential as a promising spectral tool and device applications.
A theory of the nonlinear Hall effect on a disordered lattice is a crucial step towards explorations in realistic devices, but has not been addressed. We study the nonlinear Hall response on a lattice, which allows us to introduce strong disorder numerically. We reveal a disorder-induced Berry curvature that was not discovered in the previous perturbation theories. The disorder-induced Berry curvature
induces a fluctuation of the nonlinear Hall conductivity, which anomalously increases as the Fermi energy moves from the band edges to higher energies.
More importantly, the fluctuation may explain those observations in the recent experiments. We also find signatures of localization of the nonlinear Hall effect. This work shows a territory of the nonlinear Hall effect yet to be explored.
\end{abstract}
\maketitle

{\color{blue}\emph{Introduction}.}-- The nonlinear Hall
effect behaves as a transverse Hall voltage
nonlinearly responding to a longitudinal
driving current~\cite{Sodemann15prl,Du2021NRP,Ho2021NatElec,Ma19nat,
Kang19nm,Wang2023Nature,Gao2023Science,
Lai2021NatNano}. It has attracted much attention, as a new experimental tool to reveal a number of emergent physics, such as the Berry curvature dipole~\cite{Ma19nat,Ho2021NatElec,Kang19nm},  Berry-connection polarizability, and quantum metric~\cite{WangC21PRL,Liu21PRL,Lai2021NatNano,Wang2023Nature,Gao2023Science}.
A theory of the nonlinear Hall effect on a disordered lattice is a crucial step towards explorations in realistic devices, but has not been addressed.




\begin{figure}[h!]
\centering
\includegraphics[width=\columnwidth]{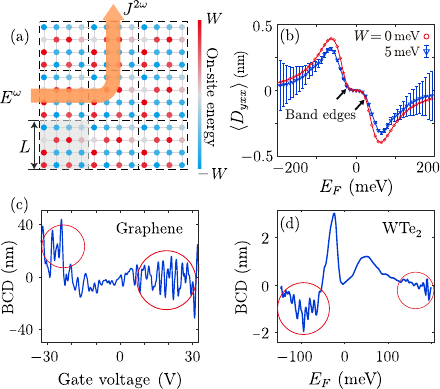}
\caption{(a) In the nonlinear Hall effect, a double-frequency transverse current $J^{2\omega}$ is induced by a low-frequency (10-1000 Hz) electric field $E^{\omega}$. Supercells (dashed boxes, $L$ for side length) allow introducing strong disorder numerically on a lattice (the lattice-site colors here show a single disorder configuration). The supercell can converge to an infinite disordered lattice within a reasonable computational power, while maintaining the lattice translational symmetry (i.e., $k_x$ and $k_y$ are still good quantum numbers). (b) The calculated nonlinear Hall effect in terms of the Berry curvature dipole $\langle D_{yxx}\rangle$ [calculated using Eq.~\eqref{Eq_BC1_2}] exhibits stronger fluctuations at higher Fermi energy $E_F$ when disorder $W\neq0$, indicated by the standard deviation bars after averaging over 5000 disorder configurations. The calculated fluctuation in (b) gives an explanation to the unexpected higher-energy stronger fluctuations of $D_{yxx}$ observed in experiments~[(c) and (d), adopted from Refs.~\cite{Ho2021NatElec} and~\cite{Ma19nat}].
}%
\label{fig_illustration}%
\end{figure}

\begin{figure}[t!]
\centering
\includegraphics[width=1\columnwidth]{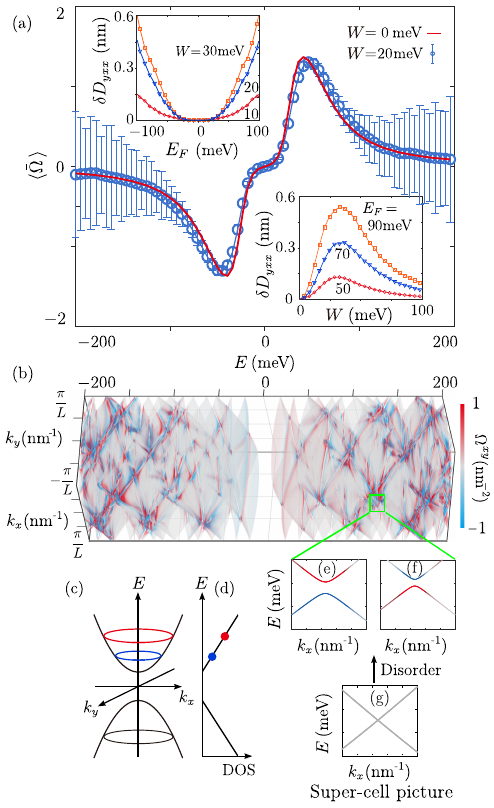}
\caption{ (a) The Berry curvature after averaging over 5000 disorder configurations $\langle \bar{\Omega}\rangle$ [calculated using Eq.~\eqref{Eq_BC}] as a function of energy $E$ for $W=0$ and $W=20$ meV. The insets show the fluctuation of the Berry curvature dipole $\delta D_{yxx}$ as functions of the Fermi energy $E_F$ and disorder strength $W$.
(b) The energy spectra $E(k_x,k_y)$ and Berry curvature $\Omega^{xy}(k_x,k_y)$ of the supercell system for a single disorder configuration ($W=20$ meV). The color scheme shows that the fluctuation of $\Omega^{xy}$ is more intense at higher energies (e.g., near $\pm$200 meV) than at low energies (e.g., near $\pm$50 meV), explaining the increasing fluctuation at higher energies in (a) and Fig.~\ref{fig_illustration}(b). (c) Schematic of the energy dispersion of the 2D Dirac model, with a density of states (DOS) linearly proportional to $E$ in (d), which explains the stronger fluctuation of the Berry curvature at higher energies in~(b).
[(e)-(g)] In the supercell picture, the degenerate states at given energies [red and blue circles in (c) and dots in (d)] turn to the band crossings in (g), which carry no Berry curvature because of violating the adiabatic conditions. Disorder can open the random gaps [e.g., positive in (e) or negative in (f)] at the crossings and generate fluctuating Berry curvature.
The supercell size $L=60$ nm in (a) and $L=8$ nm in (b). }%
\vspace{-0.5em}
\label{fig_fluctuation}%
\end{figure}

In this Letter, we study the nonlinear Hall effect on a disordered lattice~[Fig.~\ref{fig_illustration}(a)]. With the lattice treatment, we can introduce strong disorder, allowing us to explore essential topics of quantum transport, e.g., fluctuation and localization ~\cite{Anderson58PR,Lee85PRL,Lee1985RMP,Evers08RMP,Beenakker97RMP}.
Our calculations reveal
two findings in the nonlinear Hall response. (i) A fluctuation of the nonlinear Hall conductivity, which increases anomalously as the Fermi energy moves from the band edges [the black arrows in Fig.~\ref{fig_illustration}(b)] to higher energies~[the blue data in Fig.~\ref{fig_illustration}(b)]. It arises from a different mechanism of the nonlinear Hall effect, as a result of a disorder-induced Berry curvature~[Fig.~\ref{fig_fluctuation}], thus it can neither be revealed in the perturbation theories nor measured in the linear Hall conductivity.
This fluctuation may explain the recent experiments [Figs.~\ref{fig_illustration}(c) and \ref{fig_illustration}(d)], where larger nonlinear Hall conductivity fluctuations were observed at higher energies~\cite{Ma19nat,Ho2021NatElec}, but cannot be understood by the universal conductance fluctuation~\cite{Lee85PRL,Ren06PRL,Qiao08PRL} or the perturbation theories. (ii) The second feature is an ``Anderson localization", but in the nonlinear response thus is different from the previous scenarios~\cite{Lee1985RMP,Evers08RMP,Beenakker97RMP}. Our findings reveal a large territory of the nonlinear Hall effect yet to be explored.

{\color{blue}\emph{Model and the supercell method}.}-- We adopt the minimal model for the nonlinear Hall effect, i.e., the tilted 2D massive Dirac model~\cite{Du18prl},
\begin{align}
H=tk_{x}+\left(  m-\alpha
k^{2}\right)  \sigma_{z}+ v (k_{y}\sigma_{x} -  k_{x}\sigma_{y}),
\label{Model1}
\end{align}
where $(k_x, k_y)$ are the wave vectors, $k^2=k_x^2+k_y^2$, $\sigma_x, \sigma_y, \sigma_z$ are the
Pauli matrices,  $2m$ is the gap, $t$ breaks the inversion symmetry by tilting the Dirac
cone along the $x$ direction, and $\alpha$ is
introduced to regulate topological properties as $k\rightarrow \infty$.
The time reversal of the model contributes equally to the
Berry dipole, so it is enough to study this model only. Moreover, the tight-binding lattice version of the Hamiltonian is presented in Sec.~SI of Supplemental Material~\cite{Supp}.
The Hamiltonian accounts for only half of the Brillouin zone, and its time-reversal partner is in the other half of the Brillouin zone. Because of time-reversal symmetry, the Berry curvature from one half of the Brillouin zone cancels with that from the other half, so we do not have the linear Hall effect. In contrast, the Berry curvature dipole from one half of the Brillouin zone is the same as that from the other half, so we have the nonlinear Hall effect~\cite{Sodemann15prl,Du18prl}. 

The previous perturbation theories reveal that disorder plays an important role in the nonlinear Hall effect, but the
exploration was limited to weak disorder~\cite{Du19nc,Du2021NC,Joao2020RSOC,
Xiao19prb,Nandy19prb,Tiwari2021nc,Duan22prl,He2022NatNano,He2021nc}.
To deal with stronger disorder, we project the model on a 2D square lattice and introduce the Anderson disorder ~\cite{Lee1985RMP,Evers08RMP,Beenakker97RMP,Landauer1970Philosophical,Buttiker1988PRB, Fisher1981PRB,Loring2010EPL,Huang18PRB,Prodan10PRL,Prodan2011JPA,Bianco11PRB,Pozo19prl,Zhang2013CPB,Kitaev2006AP,Carcia15PRL,Weibe06RMP,Varjas20PRR}, in terms of the on-site energies uniformly distributed within
$\left[-W,W\right]$, where $W$ measures the disorder strength.
The lattice constant $a=1$ nm, $t=50$ meV nm, $v=100$ meV nm, $\alpha=100$ meV nm$^2$, and $m=40$ meV are of the same orders of those in typical massive Dirac systems~\cite{Muechle16PRX,Ma19nat,Liu2021CMS}.
The temperature is $k_BT=0.12m$.
Moreover, we adopt the supercell method to save computational power~[see Fig.~\ref{fig_illustration}(a) and Sec.~SI of Supplemental Material~\cite{Supp} for more details]. The area of the supercell is $V=L^2$, with the side lengths $L=na$, and the number of lattice sites $n^2$.

{\color{blue}\emph{Nonlinear Hall conductivity--Berry curvature dipole}.}--One of the major contributions to the nonlinear Hall conductivity (defined as a current density $j_a=\sigma_{abc}E_b E_c$ induced by two electric fields $E_b$ and $E_c$, with $a,b,c\in \{x,y,z\}$) is from the Berry curvature dipole  $\sigma_{abc}^{\text{BCD}}=(e^{3}/\hbar^{2})\tau D_{abc}$~\cite{Sodemann15prl,Low15prb}. $\tau$ is the relaxation time and the Berry curvature dipole $D_{abc}$ can be found as
\begin{equation}
D_{abc}=-\int d^2\mathbf{k} \sum_{m,p}^{E_{m}\neq E_{p}}
v_{mm}^{c}\Omega_{mp}^{ab}f_{E_{m}}%
^{\prime}, \label{Eq_BC1_2}%
\end{equation}
where the Berry curvature $\Omega
_{mp}^{ab}=-2\operatorname{Im}\left[  \mathcal{R}_{pm}%
^{a}\mathcal{R}_{mp}^{b}\right]  $,
$\mathcal{R}_{mp}^{a}=iv_{mp}^{a}/E_{mp}$,
$E_{mp}=E_{m}-E_{p},$ $v_{mp}%
^{a}=\left\langle m\right\vert \partial H/\partial k_{a}\left\vert
p\right\rangle $, $f_{E_{m}}^{\prime}=\partial f_{E_{m}}/\partial
E_{m}$, and $f$ is the Fermi function. $E_{m}$ is the eigenvalue of the $m$-th state and $\left\vert
m\right\rangle$ is the corresponding eigenstate. This integration is over the folded Brillouin zone with $k_{x/y}\in[-\pi/L,\pi/L]$. The nonlinear Hall conductivity depends also on the relaxation time $\tau$, which is irrelevant to the Berry physics and is subtracted in the experiments using the Drude conductivity~\cite{Ho2021NatElec,Ma19nat}, so we focus only on $D_{abc}$. Note that, the nonlinear Hall effect is measured at low electric-field frequencies (from $10$ to $1000$ Hz), so the energy scale and physics differ from those in the photogalvanic effects where the microwave or light frequencies are approximately $10^{9}\sim10^{14}$ Hz~\cite{Belkov2005JPCM,Drexler2013NatNano,Basken1977pssb,Sturman92book}.

Figure~\ref{fig_illustration}(b) shows the Berry curvature dipole $D_{yxx}$ of the tilted Dirac model as a
function of the Fermi energy $E_{F}$. In the absence of disorder ($W=0$), the Berry curvature and Berry curvature dipole are calculated in a unit cell or alternatively in a supercell in the weak-disorder limit (i.e., $W\rightarrow 0$). The numerical results by both approaches agree well (see Fig.~S2 of \cite{Supp}).  In the presence of disorder ($W\neq0$), we find two features in $D_{yxx}$, i.e., the {\em{localization}} and {\em{fluctuation}} effects, as manifested by the disordered-averaged Berry curvature dipole $\langle D_{yxx} \rangle$ and the corresponding fluctuation $\delta D_{yxx}$. In the numerical calculations, $\langle D_{yxx}\rangle$ is obtained after an ensemble averaging over 5000 configurations of the same disorder strength $W$. The fluctuation is defined as the standard deviation of these configurations, i.e., $\delta D_{yxx}=\sqrt{\langle D_{yxx}^2\rangle-\langle D_{yxx}\rangle^2}$.

{\color{blue}\emph{Fluctuation}.}--
As shown by the standard deviation bars in Fig.~\ref{fig_illustration}(b), the fluctuation of $\delta D_{yxx}$ increases as $E_F$ moves away from the band edges (at $E_F=\pm 40$ meV) to higher energies (e.g., $E_F=\pm 200$ meV). This can be observed more clearly in the insets of Fig.~\ref{fig_fluctuation}(a), where we show $\delta D_{yxx}$ as a function of the Fermi energy $E_F$ and disorder strength $W$. The Berry curvature dipole reaches the maximum near the band edges and decays at higher energies, but its fluctuation shows an opposite behavior [the left inset of Fig.~\ref{fig_fluctuation}(a)].
The fluctuation $\delta D_{yxx}$ can be even several times larger than the average value of $\langle D_{yxx}\rangle$ at $E_F=200$ meV.

The fluctuation of the Berry curvature dipole is attributed to the disorder-induced Berry curvature. Figure~\ref{fig_fluctuation}(a) shows the disorder-averaged Berry curvature $\langle \bar{\Omega} \rangle$ as a function of energy $E$, where $\langle ...\rangle $ means disorder average,
\begin{equation}
\bar{\Omega}(E)=-\int  d^2\mathbf{k}  \sum_{m,p}^{E_{m}\neq E_{p}} \Omega_{mp}^{xy}f_{E_{m}}%
^{\prime},
\label{Eq_BC}
\end{equation}
and the integral is for all $\mathbf{k}$ of the same energy $E$. Figure~\ref{fig_fluctuation}(a) shows that the averaged Berry curvature is stable in the absence (red data) and presence (blue data) of disorder. By contrast, its fluctuation is significantly enhanced by disorder and becomes more pronounced at higher energies.
As illustrated in Figs.~\ref{fig_fluctuation}(b)-\ref{fig_fluctuation}(g), the fluctuation of the Berry curvature is attributed to the mixing of the degenerate states of different $\mathbf{k}$.
The fluctuation is more significant at higher energies~[Fig.~\ref{fig_fluctuation}(b)] because there are more states~[Figs.~\ref{fig_fluctuation}(c) and \ref{fig_fluctuation}(d)].
Our supercell treatment also helps reveal this picture of mixed degenerate states. Within the supercell picture, the degenerate states turn to band crossings due to the Brillouin zone folding [Fig.~\ref{fig_fluctuation}(g)]. The crossings violate the adiabatic condition~\cite{Xiao2010RMP}, so at the crossings the Berry curvature from two bands is supposed to be compensated. Disorder opens random mini-gaps in these band crossings, inducing significant random fluctuations of the Berry curvature [Figs.~\ref{fig_fluctuation}(e)-\ref{fig_fluctuation}(f)]. After averaging over numerous disorder configurations, these random fluctuations in the disorder-induced Berry curvature lead to the fluctuation in the Berry curvature dipole [see Secs.~SII and SIII of Supplemental Material~\cite{Supp} for more details].


Figures~\ref{fig_illustration}(c) and \ref{fig_illustration}(d) illustrate the experimentally measured Berry curvature dipole in two distinct systems. One is the bilayer graphene~\cite{Ho2021NatElec} and the other is the bilayer WTe$_2$~\cite{Ma19nat}. In both experiments, a fluctuation of the Berry curvature dipole is observed. Remarkably, the fluctuation increases as the Fermi energy moves away from the Dirac points to higher energies~[i.e., $E_F=0$ in Fig.~\ref{fig_illustration}(d) and $V_\text{g}-V_\text{NP}=0$ in Fig.~\ref{fig_illustration}(e)]. Our theory provides a potential mechanism to understand the experimental results.

Moreover, the right inset of Fig. \ref{fig_fluctuation}(a) shows that the fluctuation increases with the disorder strength when $W<40$ meV (which is comparable to the gap of the massive Dirac model $2m$) roughly, then decreases and vanishes with further increasing disorder strength. This non-monotonic behavior can be understood by the property of the Berry curvature dipole, which first increases with the gap then drops and vanishes~\cite{Du18prl}.
The disorder-induced random mini-gaps increases with increasing disorder strength, giving rise to the non-monotonic behavior of the fluctuation with the disorder strength [Sec.~SIII B of Supplemental Material~\cite{Supp} for more details]. Therefore, the right inset of Fig. \ref{fig_fluctuation}(a) also verifies our explanation to the fluctuation of the Berry curvature dipole as a result of the disorder-induced fluctuation of the
Berry curvature.


When $W>50$ meV, we also find that $L^3\delta D_{yxx}$ remains invariant as the system size $L$ changes~[see Sec.~SIV of Supplemental Material~\cite{Supp} for more details]. This behavior differs significantly from the linear conductance fluctuations and suggests a unique scaling law in the nonlinear Hall response.

\begin{figure}[t]
\centering
\includegraphics[width=\columnwidth]{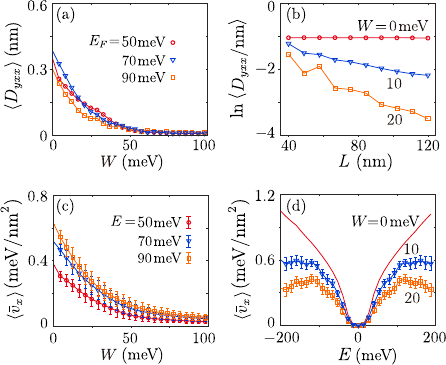}
\caption{(a) The disorder-averaged Berry curvature dipole $\langle D_{yxx}\rangle$ [calculated using Eq.~\eqref{Eq_BC1_2}] as a function of $E_F$ for different disorder strengths $W$. (b) $\ln\langle D_{yxx}\rangle$ as a function of $L$ for different $W$ at $E=-50$ meV. [(c) and (d)] The disorder-averaged velocity $\bar{v}_{x}$ [calculated using Eq.~\eqref{Eq_V}] as functions of $W$ and $E$. Here, each point is obtained averaging over 5000 disorder configurations.
To better demonstrate the fluctuation of the velocity, the standard deviation bars in (c) and (d) are magnified by 10 times.}%
\label{fig_localization}%
\vspace{-1.5em}
\end{figure}

{\color{blue}\emph{Localization}.}--
As shown in Fig.~\ref{fig_illustration}(b), the disorder-averaged Berry curvature dipole $\langle D_{yxx} \rangle$ drops as the Anderson disorder is turned on ($W\neq 0$), which can be observed more clearly in  Figs.~\ref{fig_localization}(a). Figure~\ref{fig_localization}(b) also shows that $\langle  D_{yxx}\rangle $ exhibits a nearly exponential decay with increasing supercell size $L$. This drop of the nonlinear Hall conductivity is reminiscent of the Anderson localization~\cite{Lee81PRL,Abrahams79PRL}, but the difference is that the previous Anderson localization is about the linear longitudinal conductivity. This finding of the localization of the nonlinear Hall effect has not been addressed theoretically and may be observed in future experiments.

We further show that the drop of $\langle
D_{yxx}\rangle$ has an origin similar to the Anderson localization. According to Eq.~\eqref{Eq_BC1_2}, the Berry curvature dipole is determined by the electron velocity $v$ and
Berry curvature $\Omega$ near the Fermi surface. With increasing disorder strength, the Berry curvature protected by the bulk topology is robust against disorder [Fig.~\ref{fig_fluctuation}(a)]. In contrast,
the disorder-averaged velocity
\begin{equation}
\bar{v}_{x} = \int d^2\mathbf{k} \sum_{m}
\left|v_{mm}^{x}\right|f_{E_{m}}%
^{\prime},
\label{Eq_V}
\end{equation}
decreases with increasing disorder strength [Figs.~\ref{fig_localization}(c) and \ref{fig_localization}(d)], indicating that the drop of the Berry curvature dipole has an origin similar to that of the Anderson localization.

Moreover, the fluctuation of the velocity is much smaller than that of the Berry curvature.
We need to magnify the standard deviation bars by 10 times in Figs.~\ref{fig_localization}(c) and \ref{fig_localization}(d) to show the fluctuation of the velocity. Additionally, the fluctuation of the velocity does not increase with the energy $E$, which further indicates that the fluctuation of the Berry curvature dipole in Fig. \ref{fig_illustration}(b) is mainly contributed by the fluctuation of the Berry curvature in Fig. \ref{fig_fluctuation}(a). Our calculations are performed at a higher temperature $k_BT=0.12m$ (about $55$ K). At low temperatures (around $1$ K~\cite{Droscher11PRB,Sandow01PRL}), the Coulomb gap may form in the localization regime as a result of electron-electron interactions and dramatically affect the transport ~\cite{Efros1975JPCS}.
\begin{figure}[tpb]
\centering
\includegraphics[width=1\columnwidth]{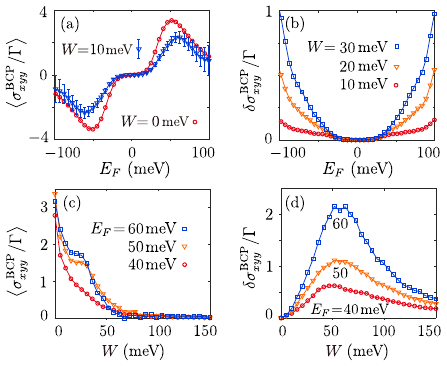}
\caption{(a) The Berry connection polarizability $\langle \sigma_{xyy}^{\mathrm{BCP}}/\Gamma \rangle$ [calculated using Eq.~\eqref{Eq_BCP}] as a function of the Fermi energy $E_F$, in the absence ($W=0$) and presence ($W=10$ meV) of disorder.
(b) The fluctuation $\delta \sigma_{xyy}^{\mathrm{BCP}}/\Gamma$ as a function of $E_F$ for different disorder strength $W$, where $\Gamma\equiv e^{3}/2\hbar\pi^{2}$. (c)  The disorder-averaged $\langle \sigma_{xyy}^{\mathrm{BCP}}/\Gamma \rangle$ and (d) the fluctuation $\delta \sigma_{xyy}^{\mathrm{BCP}}/\Gamma$ as functions of $W$ for different $E_F$. Here, the disordered data is obtained after averaging over 5000 disorder configurations. The parameters are the same as those in Figs. \ref{fig_illustration}, \ref{fig_fluctuation}, and \ref{fig_localization}.}%
\label{fig_BCP}%
\vspace{-1.5em}
\end{figure}

{\color{blue}\emph{Nonlinear Hall conductivity---Berry connection polarizability}.}--
In a $PT$-symmetric metal ($P$ for spatial inversion and $T$ for time-reversal),
the nonlinear Hall effect can also emerge as a result of the Berry connection polarizability~\cite{WangC21PRL,Liu21PRL}, which measures the distance between quantum states and deflects electronic carriers to the perpendicular direction.
We show that the fluctuation and localization also present in the Berry connection polarizability under strong disorder.

The Berry connection polarizability can be found as%
\begin{equation}
\sigma_{abc}^{\mathrm{BCP}}=\int d^2\mathbf{k}\Gamma\sum_{m,p}^{E_{m}\neq E_{p}}\left(  \frac{\mathcal{G}_{mp}^{bc}%
v_{mm}^{a}-\mathcal{G}_{mp}^{ac}v_{mm}^{b}%
}{E_{mp}}\right)  f_{E_{m}}^{\prime},
\label{Eq_BCP}
\end{equation}
where $\Gamma=e^{3}/2\hbar\pi^{2}$ and $\mathcal{G}_{mp}^{bc
}=\operatorname{Re}\mathcal{R}_{pm}^{b}\mathcal{R}_{mp}%
^{c}$.
To have the $PT$-symmetry, we consider a four-band tilted Dirac
model~\cite{WangC21PRL,Liu21PRL,WangY2022DarXiv}
\begin{align}
H'=tk_{x}+\left(  m-\alpha
k^{2}\right)  \tau_{z}+ v k_{x}\tau_{x}+ v k_{y}\tau_y\sigma_{x},
\label{Model2}
\end{align}
which obeys $PT H'\left(\mathbf{k}\right)(PT)^{-1}=H'\left(\mathbf{k}\right)$, where the $PT$-symmetry operator  $PT=-i\sigma_{y} K$ and $K$ means the complex conjugate. The parameters are the same as those in Eq. (\ref{Model1}).

Figure~\ref{fig_BCP} shows the results for the Berry connection polarizability $\sigma_{xyy}^{\mathrm{BCP}}$. In the presence of disorder, the Berry connection polarizability also shows the similar fluctuation and localization, i.e., the exponential decay with increasing system size and significant Fermi-energy-dependent fluctuations. In Sec.~SV of Supplemental Material~\cite{Supp}, we provide more numerical results for the Berry connection polarizability.

This similarity also verifies our explanation to the fluctuation and localization of the Berry curvature dipole, because the Berry connection polarizability and Berry curvature are related as the real and imaginary parts
of the quantum geometry tensor~\cite{Provost1980ComMP},
\begin{align}
T^{ab}=\mathcal{G}^{ab}-i\Omega^{ab}/2,
\label{Eq:Metric}
\end{align}
where $\Omega
_{mp}^{ab }=-2\operatorname{Im}\left[  \mathcal{R}_{pm}%
^{a }\mathcal{R}_{mp}^{b }\right]  $ is the Berry curvature in Eq. (\ref{Eq_BC}), $\mathcal{G}_{mp}^{ab
}=\operatorname{Re}\mathcal{R}_{pm}^{a }\mathcal{R}_{mp}%
^{b }$ is the quantum metric in Eq. (\ref{Eq_BCP}), and $\mathcal{R}_{mp}^{a }=i\left\langle m\right\vert \partial H/\partial k_{a}\left\vert
p\right\rangle/(E_{m} -E_{p} )$.
Therefore, both the Berry curvature dipole and Berry connection polarizability are supposed to show the similar fluctuation and localization.

\begin{acknowledgments}
This work was supported by the National Key R$\&$D Program of China (2022YFA1403700), the Innovation Program for Quantum Science and Technology (2021ZD0302400), the National Natural Science Foundation of China (11925402 and 12350402), Guangdong province (2020KCXTD001 and 2016ZT06D348), the Science, Technology and Innovation Commission of Shenzhen Municipality (ZDSYS20170303165926217, JAY20170412152620376, and KYTDPT20181011104202253). Rui Chen thanks helpful discussions with Bo Fu and acknowledges the support of the National Natural Science Foundation of China (12304195) and the Chutian Scholars Program in Hubei Province. The numerical calculations were supported by Center for Computational Science and Engineering of SUSTech.
\end{acknowledgments}

\bibliographystyle{apsrev4-1-etal-title_6authors}
\bibliography{refs-transport}

\end{document}